\documentclass[runningheads]{svmult}

\usepackage{makeidx}   
\usepackage{graphicx}  
\usepackage{subeqnar}  
\usepackage{multicol}  
\usepackage{physprbb}  
\makeindex             

\begin{document}
\title*{Quantum Phase Transitions}
%
%
\author{Thomas Vojta}
\institute{Institut f\"ur Physik, Technische Universit\"at Chemnitz,
           D-09107 Chemnitz, Germany}

\maketitle              

\begin{abstract}
Phase transitions which occur at zero temperature when some non-thermal
parameter like pressure, chemical composition or magnetic field is changed
are called quantum phase transitions.
They are caused by quantum fluctuations which are a consequence of
Heisenberg's uncertainty principle. These lecture notes
give a pedagogical introduction
to quantum phase transitions. After collecting a few basic facts about
phase transitions and critical behavior we discuss the importance
of quantum mechanics and the relation between
quantum and classical transitions as well as their experimental relevance.
As a primary example we then consider the Ising model in a transverse field.
We also briefly discuss quantum phase transitions in itinerant
electron systems and their connection to non-Fermi liquid behavior.
\end{abstract}

\section{Introduction: From the melting of ice to quantum criticality}

When a piece of ice is taken out of the freezer, in the beginning
its properties change only slowly with increasing temperature.
At $0^\circ$C, however, a
drastic change happens. The thermal motion of the water molecules becomes
so strong that it destroys the crystal structure. The ice melts, and a new
phase of water forms, the liquid phase. This process is an example for a
phase transition.
At the transition temperature of $0^\circ$C the solid
and the liquid phases of water coexist.
A finite amount of heat, the so-called latent heat, is necessary to transform
the ice into liquid water. Phase transitions which involve latent heat
are usually called first-order transitions.

Another well known example of a phase transition is the magnetic
transition of iron. At room temperature iron is ferromagnetic, i.e.\
it possesses a spontaneous magnetization. With
increasing temperature the magnetization decreases continuously.
It vanishes at the Curie temperature of
$770^\circ$C, and above this temperature iron is paramagnetic.
This is a phase transition from a ferro- to a paramagnet. In contrast to the
previous example there is no phase coexistence at the transition
temperature, the two phases rather become indistinguishable.
Consequently, there is no latent heat.
This type of phase transitions are called continuous transitions
or second-order transitions. They are the result of a competition
between order and thermal fluctuations. When approaching the
transition point the typical length and time scales of the fluctuations diverge,
leading to singularities in many physical quantities.
The diverging correlation length at a continuous
phase transition was first observed in 1869 in a
famous experiment by Andrews \cite{Andrews1869}.  He
discovered that the properties of the liquid and the
vapor phases of carbon dioxide became indistinguishable at a temperature
of about 31\,$^\circ$C and 73 atmospheres pressure. In the vicinity of
this point the carbon dioxide became opaque, i.e. it strongly scattered
visible light, indicating that the length
scale of the density fluctuations had reached the wave length of the light.
Andrews called this special point in the phase diagram
the critical point and the strong light scattering in its vicinity
the critical opalescence.

More formally one can define a phase transition as the occurrence of a
singularity in the thermodynamic quantities as functions of the external
parameters. Phase transitions have played, and continue to play,
an essential role in shaping our world. The large scale
structure of the universe is the result of a sequence of
phase transitions during the very early stages
of its development. Even our everyday life is unimaginable without the
never ending transformations of water between ice, liquid and
vapor.
Understanding the properties of phase transitions and in
particular those of critical points has been a great
challenge for theoretical physics. More than a century has gone by from
the first discoveries
until a consistent picture emerged. However, the theoretical concepts
established during this development, viz., scaling and the
renormalization group \cite{Wilson71},
now belong to the central paradigms of modern physics.

In the last decade considerable attention has concentrated on a class of
phase transitions which are qualitatively very different from
the examples discussed above. These new transitions occur at zero
temperature when a non-thermal parameter like pressure, chemical composition
or magnetic field is changed. The fluctuations which destroy the long-range
order in these transitions cannot be of thermal nature since thermal
fluctuations do not exist at zero temperature.
Instead, they are quantum fluctuations which are a consequence of Heisenberg's
uncertainty principle. For this reason phase transitions at zero temperature
are called quantum phase transitions, in contrast to thermal or classical
phase transitions at finite temperatures.
(The justification for calling all thermal phase transitions classical will
become clear in Sec.\ \ref{sec:qm})

\begin{figure}[t]
  \centerline{\includegraphics*[width=7.5cm]{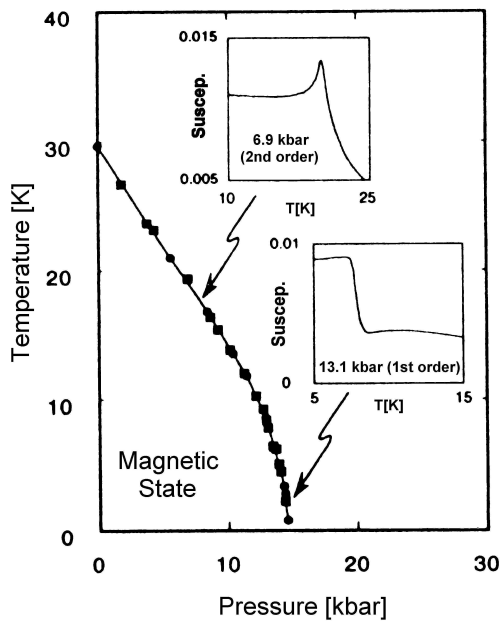}}
  \caption{Magnetic phase diagram of MnSi (after Ref.\
   \protect\cite{PML95}). The peculiar behavior of
   the magnetic susceptibility shown in the insets is a consequence of the
   interplay between the magnetic and other fermionic degrees of freedom
   \protect\cite{us_fm}).}
  \label{fig:MnSi}
\end{figure}

As an illustration of classical and quantum phase transitions we show
the magnetic phase diagram of the transition metal compound MnSi in Fig. \ref{fig:MnSi}.
At ambient pressure MnSi is a paramagnetic metal
for temperatures larger than $T_c=30$\,K. Below $T_c$ it orders
ferromagnetically\footnote{Actually, the ordered state is a spin spiral in
the (111) direction of the crystal. Its wavelength is very long (200 \AA)
so that the material behaves like a ferromagnet for most purposes.}
but remains metallic. This transition is a {\em thermal} continuous phase transition
analogous to that in iron discussed above.
Applying pressure reduces the transition temperature, and at about 14 kBar the
magnetic phase vanishes. Thus, at 14 kBar MnSi undergoes a quantum phase
transition from a ferro- to a paramagnet. While this is a very obvious
quantum phase transition, its properties are rather complex due to the
interplay between the magnetic degrees of freedom and other fermionic
excitations (for details see \cite{us_fm}).

At a first glance quantum phase transitions seem to be a purely academic
problem since they occur at isolated parameter values and at zero temperature
which is inaccessible in an experiment. However, within the last decade
it has turned out that the
opposite is true. Quantum phase transitions do have important,
experimentally relevant
consequences, and they are believed to provide keys to many new and exciting
phenomena in condensed matter physics, such as the quantum Hall effects,
the localization problem, non-Fermi liquid behavior in metals or high-$T_c$
superconductivity.

These lecture notes are intended as a pedagogical introduction
to quantum phase transitions. In section \ref{sec:concepts} we first collect
a few basic facts about
phase transitions and critical behavior.
In section \ref{sec:qm} we then investigate the importance of quantum mechanics
for the physics of phase transitions and
the relation between
quantum and classical transitions.
Section \ref{sec:qcp} is devoted to a detailed analysis of the
physics in the vicinity of a quantum critical point.
We then discuss two examples: In section \ref{sec:ising} we consider
one of the paradigmatic models in this field, the Ising model in a
transverse field, and in section \ref{sec:non-fermi}
we briefly discuss quantum phase transitions in itinerant
electron systems and their connection to non-Fermi liquid behavior.

Those readers who want to learn more details about quantum phase transitions
and the methods used to study them, should consult one of the recent
review articles, e.g. Refs.\
\cite{sgcs97,kb97,vojta2000},
or the excellent text book on quantum phase transitions
by Sachdev \cite{sachdev00}.

\section{Basic concepts of phase transitions and critical behavior}
\label{sec:concepts}
In this section we briefly collect the basic concepts of continuous phase
transitions and critical behavior which are necessary for the
later discussions. For a detailed exposure the reader is referred
to one of the text books on this subject,
e.g., those by Ma \cite{Ma76} or  Goldenfeld \cite{Goldenfeld92}).

A continuous phase transition can usually be characterized by an
order parameter, a concept first introduced by Landau \cite{Landau37}.
An order parameter is a thermodynamic quantity that is zero in one phase
(the disordered) and non-zero and non-unique in the other (the ordered)
phase. Very often the choice of an order parameter for a particular
transition is obvious as, e.g., for the ferromagnetic transition where
the total magnetization is an order parameter. Sometimes, however,
finding an appropriate order parameter is a complicated problem by
itself, e.g., for the disorder-driven localization-delocalization transition
of non-interacting electrons.

\begin{figure}[t]
\centerline{\includegraphics*[width=13cm]{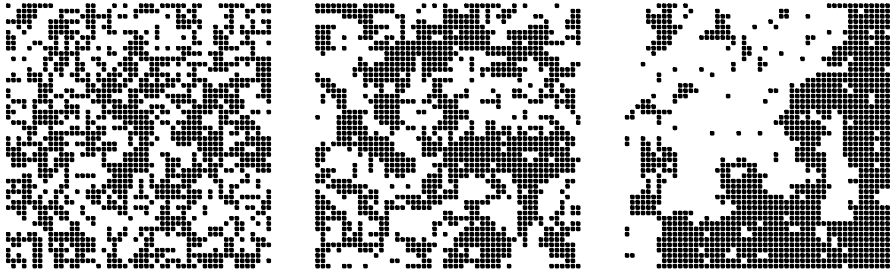}}
\caption{Snapshots of the spin configuration of a two-dimensional
   Ising model at different temperatures (a black dot corresponds to
   spin up, an empty site means spin down). From left to right:
   $T=2T_c$, $T=1.3 T_c$, $T=T_c$. The correlation length increases
   with decreasing temperature and diverges at $T=T_c$. Here the system
   fluctuates at all length scales.}
\label{fig:conf}
\end{figure}

While the thermodynamic average of the order parameter is zero in the
disordered phase, its fluctuations are non-zero. If the phase transition
point, i.e., the critical point, is approached the spatial
correlations of the
order parameter fluctuations become long-ranged. Close to the
critical point their typical length scale, the correlation length $\xi$,
diverges as
\begin{equation}
  \xi \propto |t|^{-\nu}
\end{equation}
where $\nu$ is the correlation length critical exponent and
$t$ is some dimensionless measure of the distance from the critical point.
If the transition occurs at a non-zero temperature $T_c$,
it can be defined as $t=|T-T_c|/T_c$.
The divergence of the correlation length when approaching the transition is
illustrated in Fig.\ \ref{fig:conf} which shows computer
simulation results for the phase transition in a two-dimensional
Ising model.

In addition to the long-range correlations in space there are
analogous long-range correlations of the order parameter
fluctuations in time. The typical time scale for a decay of
the fluctuations is the correlation (or equilibration) time
$\tau_c$. As the critical point is approached the correlation
time diverges as
\begin{equation}
  \tau_c \propto \xi^z \propto |t|^{-\nu z}
\label{eq:correlation time}
\end{equation}
where $z$ is the dynamical critical exponent.
Close to the critical point there is no characteristic
length scale other than $\xi$ and no characteristic time
scale other than $\tau_c$.\footnote{Note that a microscopic
cutoff scale must be present to explain non-trivial critical
behavior, for details see, e.g., Goldenfeld \cite{Goldenfeld92}.
In a solid such a scale is, e.g., the lattice
spacing.}
As already noted by Kadanoff \cite{Kadanoff66}, this is the
physics behind Widom's scaling hypothesis \cite{Widom65},
which we will now discuss.

Let us consider a classical system, characterized by its Hamiltonian
\begin{equation}
  H(p_i,q_i)= H_{kin}(p_i)+ H_{pot}(q_i)
\end{equation}
where $q_i$ and $p_i$ are the generalized coordinates and momenta,
and $H_{kin}$ and $H_{pot}$ are the kinetic and potential energies,
respectively.\footnote{Velocity dependent potentials like in the case
of charged particles in an electromagnetic field are excluded.}
In such a system statics and dynamics decouple, i.e., the momentum
and position sums in the partition function
\begin{equation}
  Z= \int \prod dp_i e^{-H_{kin}/k_B T} ~\int \prod dq_i e^{-H_{pot}/k_B T}
   = Z_{kin}  Z_{pot}
\label{eq:classical Z}
\end{equation}
are completely independent from each other.
The kinetic contribution to the free energy
density $f=-(k_B T/V) \log Z$ will usually not display any singularities,
since it derives from the product of simple Gaussian integrals. Therefore
one can study phase transitions and the critical behavior using effective
time-independent
theories like the classical Landau-Ginzburg-Wilson theory. In this type of
theories
the free energy is expressed as a functional of the order parameter
$M({\bf r})$, which is time-independent but fluctuates in space.
All other degrees of freedom have been integrated
out in the derivation of the theory starting from a microscopic
Hamiltonian. In its simplest form \cite{Wilson71,Landau37,Ginzburg60}
valid, e.g.,
for an Ising ferromagnet, the Landau-Ginzburg-Wilson functional $\Phi[M]$ reads
\begin{eqnarray}
  \Phi[M] &=& \int d^d r ~M({\bf r})
    \left (-\frac {\partial^2}{\partial {\bf r}^2} + t  \right ) M({\bf r})
    + u \int d^d r ~ M^4({\bf r}) - B \int d^d r ~ M({\bf r}), \nonumber \\
      Z &=& \int D[M] e^{-\Phi[M]} \quad ,
\label{eq:classical LGW}
\end{eqnarray}
where $B$ is the field conjugate to the order parameter (the magnetic field
in case of a ferromagnet).

Since close to the critical point the correlation length is the only relevant
length scale, the physical properties must be unchanged, if
we rescale all lengths in the system by a common factor $b$,
and at the same time adjust the external parameters in such a way
that the correlation length retains its old value. This observation
gives rise to
the homogeneity relation for the free energy density,
\begin{equation}
 f(t,B) = b^{-d} f(t\, b^{1/\nu}, B\, b^{y_B}).
\label{eq:widom}
\end{equation}
Here $y_B$ is another critical exponent and $d$ is the space dimensionality.
The scale factor $b$ is an arbitrary positive number.
Analogous homogeneity relations for other thermodynamic quantities
can be obtained by differentiating the free energy. The homogeneity law
(\ref{eq:widom}) was first obtained phenomenologically by
Widom \cite{Widom65}. Within the framework of the renormalization group
theory \cite{Wilson71} it can be derived from first principles.

In addition to the critical exponents $\nu,~ y_B$ and $z$ defined above,
a number of other exponents is in common use. They describe the
dependence of the order parameter and its correlations on the
distance from the
critical point and on the field conjugate to the order parameter.
The definitions of the most commonly used
critical exponents are summarized in Table \ref{table:exponents}.
\begin{table}[t]
\caption{Definitions of the commonly used critical exponents in
  the `magnetic language', i.e., the order parameter is the magnetization
  $m=\langle M \rangle$, and the conjugate field is a magnetic field $B$. $t$ denotes the
  distance from the critical point and $d$ is the space dimensionality.
  (The exponent $y_B$ defined in (\ref{eq:widom}) is related to $\delta$
  by $y_B=d \, \delta /(1+\delta)$.)}
\renewcommand{\arraystretch}{1.2}
\vspace{2mm}
\begin{tabular*}{\textwidth}{c@{\extracolsep\fill}ccc}
\hline
&exponent& definition & conditions \\
\hline
specific heat &$\alpha$& $c \propto |t|^{-\alpha}$ & $t \to 0, B=0$\\
order parameter& $\beta$ & $m \propto (-t)^\beta$ & $t \to 0$ from below, $B=0$\\
susceptibility& $\gamma$ & $\chi \propto |t|^{-\gamma}$ & $t \to 0, B=0$\\
critical isotherm & $\delta$ & $B \propto |m|^\delta {\rm sign}(m)$ & $B \to 0, t=0$\\
\hline
correlation length& $\nu$ & $\xi \propto |t|^{-\nu}$ & $t \to 0, B=0$\\
correlation function& $\eta$ & $G(r) \propto |r|^{-d+2-\eta}$ & $t=0, B=0$\\
\hline
dynamical& $z$ & $\tau_c \propto \xi^{z}$ & $t \to 0, B=0$\\
\hline
\end{tabular*}
\vspace*{2mm}
\label{table:exponents}
\end{table}
Note that not all the exponents defined in Table \ref{table:exponents}
are independent from each other.
The four thermodynamic exponents $\alpha, \beta,\gamma,\delta$ can
all be obtained from the free energy (\ref{eq:widom}) which contains
only two independent exponents.
They are therefore connected by the so-called scaling relations
\begin{eqnarray}
2- \alpha =  2 \beta +\gamma~, \qquad
2 - \alpha = \beta ( \delta + 1)~.
\end{eqnarray}
Analogously, the exponents of the correlation length and correlation
function are connected by two so-called hyperscaling relations
\begin{eqnarray}
2- \alpha =  d\,\nu~,  \qquad
\gamma = (2-\eta) \nu~.
\end{eqnarray}
Since statics and dynamics decouple in classical statistics the dynamical
exponent $z$ is completely independent from all the others.

The set of critical exponents completely characterizes the critical behavior
at a particular phase transition. One of the most
remarkable features of continuous phase transitions is universality, i.e.,
the fact that the critical exponents are the same for entire classes of
phase transitions which may occur in very different physical systems.
These classes, the so-called universality classes, are determined only
by the symmetries of the Hamiltonian and the spatial dimensionality
of the system. This implies that the critical exponents of a
phase transition occurring in nature can be determined exactly
(at least in principle) by investigating
any simplistic model system belonging to the same universality class,
a fact that makes the field very attractive for theoretical physicists.
The mechanism behind universality is again the divergence of the correlation
length. Close to the critical point the system effectively averages over
large volumes rendering the microscopic details of the Hamiltonian
unimportant.

The critical behavior at a particular transition is crucially determined
by the relevance or irrelevance of order parameter fluctuations.
It turns out that fluctuations become increasingly important if the
spatial dimensionality of the system is reduced. Above a certain
dimension, called the upper critical dimension $d_c^+$, fluctuations are
irrelevant, and the critical behavior is identical to that predicted by
mean-field theory (for systems with short-range interactions and a scalar
or vector order parameter $d_c^+=4$). Between $d_c^+$ and a second special dimension, called
the lower critical dimension $d_c^-$, a phase transition still exists but
the critical behavior is different from mean-field theory. Below $d_c^-$
fluctuations become so strong that they completely suppress the ordered phase.

\section{How important is quantum mechanics?}
\label{sec:qm}

The question of to what extent quantum mechanics is important
for understanding a continuous phase transition is a multi-layered
question. One may ask, e.g., whether quantum mechanics is necessary
to explain the existence and the properties of the ordered phase.
This question can only be decided on a case-by-case basis, and
very often quantum mechanics is essential as, e.g., for superconductors.
A different question to ask would be how important quantum mechanics
is for the asymptotic behavior close to the critical point and thus for
the determination of the universality class the transition belongs to.

It turns out that the latter question has a remarkably clear and
simple answer: Quantum mechanics does {\em not} play any role for
the critical behavior if the transition occurs at a finite temperature.
It does play a role, however, at zero temperature. In the following
we will first give a simple argument explaining these facts.
To do so it is useful to distinguish fluctuations with predominantly thermal
and quantum character depending on whether their thermal energy $k_B T$ is
larger or smaller than the quantum energy scale $\hbar \omega_c$.
We have seen in the preceeding section that the typical time scale $\tau_c$
of the fluctuations diverges as a continuous transition is approached.
Correspondingly, the typical frequency scale $\omega_c$ goes to zero and with it the typical
energy scale
\begin{equation}
  \hbar \omega_c \propto |t|^{\nu z}~.
  \label{eq:energy scale}
\end{equation}
Quantum fluctuations will be important as long as this typical energy scale
is larger than  the thermal energy $k_B T$.
If the transition occurs
at some finite temperature $T_c$ quantum mechanics will therefore become
unimportant for $|t|<t_x$ with the crossover distance $t_x$ given by
\begin{equation}
 t_x \propto T_c^{1/\nu z}~.
 \label{eq:crossover t}
\end{equation}
We thus find that the
critical behavior asymptotically close to the transition is entirely
classical if the transition temperature $T_c$ is nonzero. This justifies
to call all finite-temperature phase transitions classical transitions,
even if the properties of the ordered state are completely determined
by quantum mechanics as is the case, e.g., for the superconducting
phase transition of mercury at $T_c=4.2$ K.
In these cases quantum fluctuations are obviously
important on microscopic scales, while classical thermal fluctuations
dominate on the macroscopic scales that control the critical behavior.
This also implies that only universal quantities like the critical
exponents will be independent of quantum mechanics while
non-universal quantities like the critical temperature
in general will depend on quantum mechanics.

If, however, the transition occurs at zero temperature as a function of
a non-thermal parameter like the pressure $p$, the crossover distance $t_x$
equals zero since there are no thermal fluctuations.
(Note that at zero temperature the distance $t$ from the critical point
cannot be defined via the reduced temperature. Instead, one can define
$t=|p-p_c|/p_c$.)
Thus, at zero temperature the condition
$|t|<t_x$ is never fulfilled, and quantum mechanics
will be important for the critical behavior. Consequently,
transitions at zero temperature are called quantum phase transitions.

Let us now generalize the homogeneity law (\ref{eq:widom}) to the
case of a quantum phase transition. We consider a system
characterized by a Hamiltonian $H$.
In a quantum problem kinetic and potential part of $H$
in general do not commute. In contrast to the classical partition
function (\ref{eq:classical Z}) the quantum mechanical partition function
does {\em not} factorize, i.e., statics and dynamics are always coupled.
The canonical density operator $e^{-H/k_B T}$ looks
exactly like a time evolution operator in imaginary time
$\tau$ if one identifies
\begin{equation}
1/k_B T = \tau = -i\Theta /\hbar
\label{eq:imaginary time}
\end{equation}
where $\Theta$ denotes the real time. This naturally
leads to the introduction of an imaginary time direction into the
system. An order parameter field theory analogous to
the classical Landau-Ginzburg-Wilson theory (\ref{eq:classical LGW})
therefore needs to be formulated in terms of space and time dependent fields.
The simplest example of a quantum Landau-Ginzburg-Wilson functional,
valid for, e.g., an Ising model in a transverse field (see
section \ref{sec:ising}), reads
\begin{eqnarray}
  \Phi[M] &=& \int_0^{1/k_B T} d\tau \int d^d r ~M({\bf r},\tau)
    \left (-\frac {\partial^2}{\partial {\bf r}^2}
    -\frac {\partial^2}{\partial \tau^2}+ t  \right ) M({\bf r},\tau)
    ~+ \nonumber\\
    &+& u \int_0^{1/k_B T} d\tau \int d^d r ~ M^4({\bf r},\tau)
    ~-~ B \int_0^{1/k_B T} d\tau \int d^d r ~ M({\bf r},\tau)~.
      \label{eq:quantum LGW}
\end{eqnarray}
Let us note that the coupling of statics and dynamics
in quantum statistical mechanics also leads to the fact that the
universality classes for quantum phase transitions are smaller
than those for classical transitions. Systems which belong
to the same classical universality class may display
different quantum critical behavior, if their dynamics differ.

The classical homogeneity law (\ref{eq:widom}) for the free energy density can
now easily be adopted to the case of a quantum phase transition.
At zero temperature the imaginary time acts similarly to an additional
spatial dimension since the extension of the system in this direction
is infinite. According to (\ref{eq:correlation time}), time scales like
the $z$th power of a length. (In the simple example
(\ref{eq:quantum LGW}) space and time enter the
theory symmetrically leading to $z=1$.)
Therefore, the homogeneity law for the free energy density at
zero temperature reads
\begin{equation}
 f(t,B) = b^{-(d+z)} f(t\, b^{1/\nu},B\, b^{y_B})~.
\label{eq:quantum widom}
\end{equation}
Comparing this relation to the classical homogeneity law (\ref{eq:widom})
directly shows that a quantum phase transition in $d$ spatial dimensions is
equivalent to a classical transition in $d+z$ spatial dimensions.
Thus, for a quantum phase transition
the upper critical dimension, above which
mean-field critical behavior becomes exact, is reduced by $z$
compared to the corresponding classical transition.
Note, however, that the mapping of a quantum phase transition to the
equivalent classical transition in general leads to unusual
anisotropic classical systems.
Furthermore, the mapping is valid for the thermodynamics only.
Other properties like
the real time dynamics at finite temperatures require more careful
considerations (see, e.g., Ref.\ \cite{sachdev00}).

Now the attentive reader may again ask: Why are quantum phase transitions more than an academic
problem? Any experiment is done at a non-zero temperature where,
as we have explained above, the asymptotic critical behavior is classical.
The answer is provided by the crossover condition
(\ref{eq:crossover t}): If the transition
temperature $T_c$ is very small quantum fluctuations will remain important
down to very small $t$, i.e., very close to the phase boundary.
At a more technical level, the behavior at small but non-zero temperatures
is determined by the crossover between two types of critical behavior,
viz. quantum critical behavior at $T=0$ and classical critical behavior
at non-zero temperatures. Since the `extension of the system in
imaginary time direction' is given by the inverse temperature $1/k_B T$ the
corresponding crossover scaling is equivalent to finite size scaling in
imaginary time direction. The crossover from quantum to classical behavior will
occur when the correlation time $\tau_c$ reaches $1/k_B T$ which is equivalent
to the condition (\ref{eq:crossover t}).
By adding the temperature as an explicit parameter and taking into
account that in imaginary-time formalism it scales like an
inverse time (\ref{eq:imaginary time}), we can generalize the quantum
homogeneity law (\ref{eq:quantum widom}) to finite temperatures,
\begin{equation}
 f(t,B,T) = b^{-(d+z)} f(t\, b^{1/\nu},B\, b^{y_B}, T \, b^z)~.
\label{eq:temperature quantum widom}
\end{equation}
Once the critical exponents $z$, $\nu$, and $y_B$ and the scaling function
$f$ are known this relation completely
determines the thermodynamic properties close to the quantum phase transition.

\section{Quantum-critical points}
\label{sec:qcp}
\begin{figure}[t]
  \centerline{\includegraphics*[width=6.1cm]{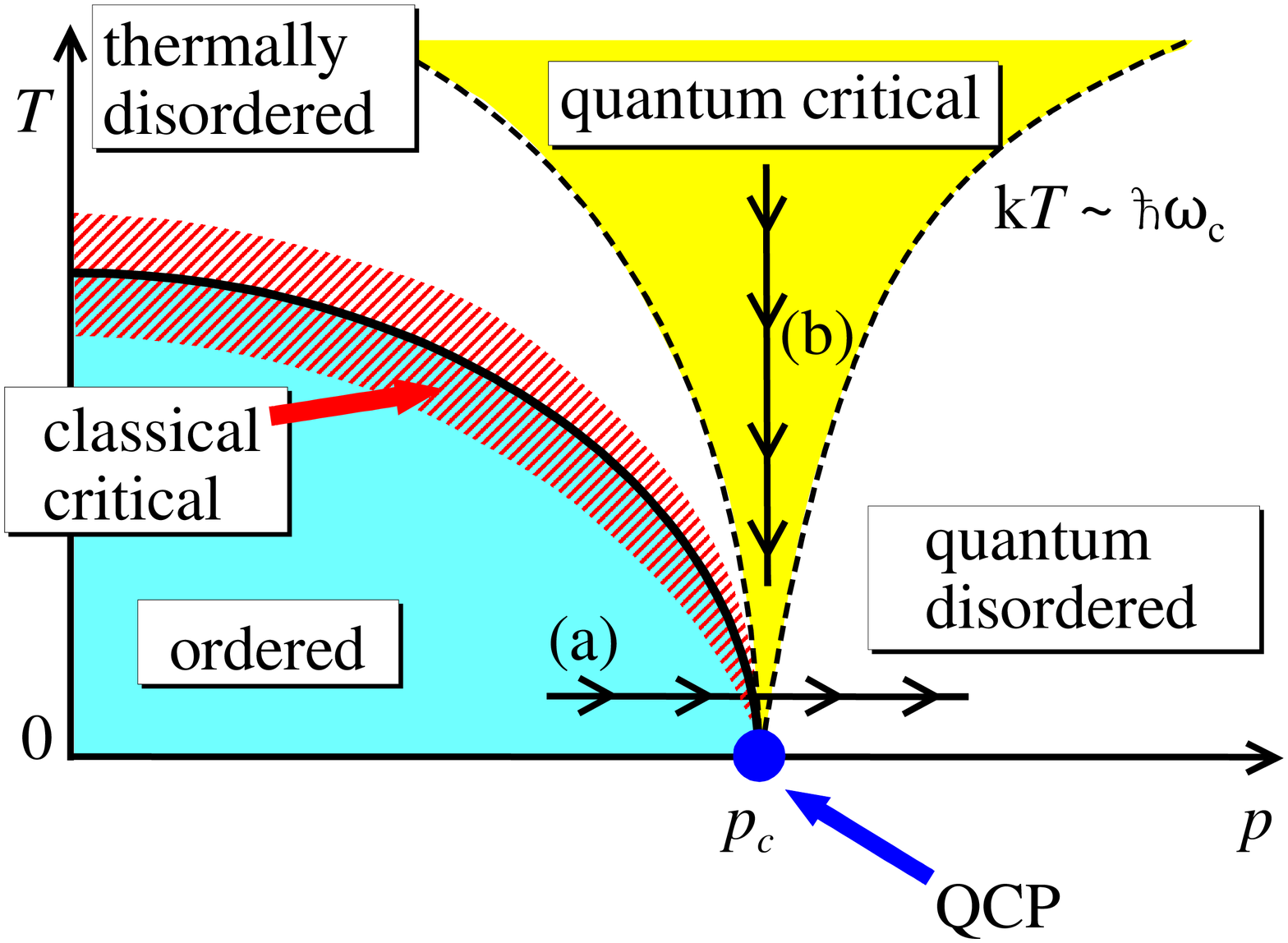}
              \includegraphics*[width=6.1cm]{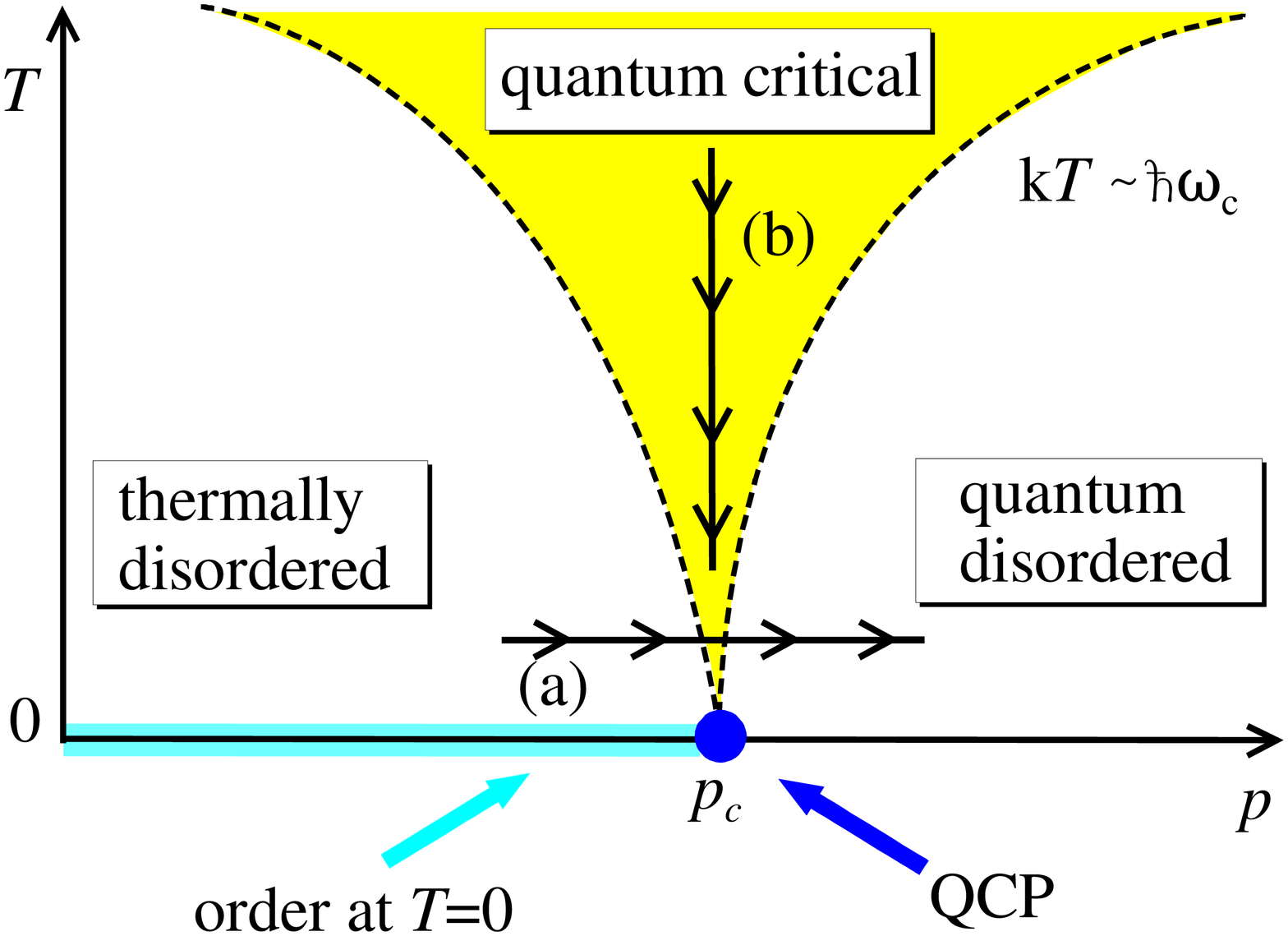}}
  \caption{Schematic phase diagrams in the vicinity of a quantum
     critical point (QCP) for situations where an ordered phase
     exists at finite temperatures (left) and situations where order
     exists at zero temperature only (right).
     The solid line marks the boundary
     between ordered and disordered phase. The different regions
     and paths (a) and (b) are discussed
     in the text.
     }
\label{fig:schematic phase diagram}
\end{figure}

We now use the general scaling picture developed in the last section
to discuss the physics in the vicinity of the quantum critical
point. There are two qualitatively different types of phase diagrams
depending on the existence or non-existence of long-range order at finite
temperatures. These phase diagrams are schematically shown in Fig.\
\ref{fig:schematic phase diagram}.
Here $p$ stands for the (non-thermal) parameter which tunes the quantum phase
transition. In addition to the phase boundary the phase diagrams
show a number of crossover lines where the properties of the
system change smoothly. They separate regions with different characters
of the fluctuations.

The first type of phase diagrams describes situations where an
ordered phase exists at finite temperatures.
As discussed in the last section classical fluctuations will
dominate in the vicinity of the phase boundary (in the hatched region in
Fig.\ \ref{fig:schematic phase diagram}). According to (\ref{eq:crossover t})
this region becomes narrower with decreasing temperature.
An experiment performed along path (a) will therefore observe a
crossover from quantum critical behavior away from the transition
to classical critical behavior asymptotically close to it. At very low
temperatures the classical region may become so narrow that it is actually
unobservable in an experiment.
In the quantum disordered region ($p>p_c$, small $T$) the physics is dominated
by quantum fluctuations, the system essentially looks as in its quantum disordered
ground state at $p>p_c$. In contrast,
in the thermally disordered region the long-range order is destroyed mainly
by thermal fluctuations.

Between the quantum disordered and the thermally disordered regions
is the so-called quantum critical region \cite{CHN89}, where both types
of fluctuations are important.
It is located at the critical $p$ but, somewhat counter-intuitively, at
comparatively high temperatures. Its boundaries are also determined by
(\ref{eq:crossover t}) but in general with a prefactor different from that
of the asymptotic classical region.
The physics in the quantum critical region is controlled by the quantum
critical point:
The system 'looks critical' with respect to $p$ (due to quantum
fluctuations) but is driven away from
criticality by thermal fluctuations (i.e., the critical singularities are
exclusively protected by the temperature $T$).
In an experiment carried out along path (b) the physics will therefore
be dominated by the critical fluctuations which diverge according to the
temperature scaling at the quantum critical point.

The second type of phase diagram occurs
if an ordered phase exists at zero temperature only (as is the case for
two-dimensional quantum antiferromagnets). In this case there will be no
true phase transition in any experiment. However, an experiment along
path (a) will show a very sharp crossover which becomes more pronounced with
decreasing temperature. Furthermore, the system will display
quantum critical behavior in the above-mentioned quantum critical region
close to the critical $p$ and at higher temperatures.

\section{Example: Transverse field Ising model}
\label{sec:ising}
In this section we want to illustrate the general ideas presented
in sections \ref{sec:qm} and \ref{sec:qcp} by discussing a paradigmatic
example, viz. the Ising model in a transverse field.
An experimental realization of this model can be found in the
low-temperature magnetic properties of
LiHoF$_4$. This material is an ionic crystal, and at sufficiently
low temperatures the only magnetic degrees of freedom are the spins
of the Holmium atoms. They have an easy axis, i.e. they prefer to point
up or down with respect to a certain crystal axis. Therefore they can
be represented by Ising spin variables. Spins at different
Holmium atoms interact via a magnetic dipole-dipole interaction.
Without external magnetic field the ground state is a fully polarized
ferromagnet.\footnote{In the case of the dipole-dipole interaction
the ground state configuration depends on the geometry of the lattice.
In LiHoF$_4$ it turns out to be ferromagnetic.}

\begin{figure}[t]
  \centerline{\includegraphics[width=9cm]{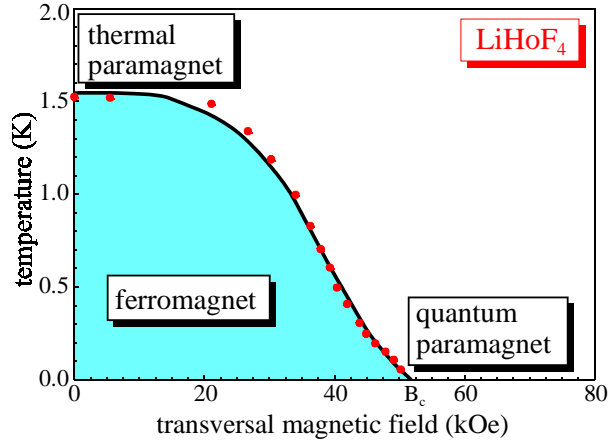}}
  \caption{Magnetic phase diagram of LiHoF$_4$ (after \protect\cite{BRA96}).}
\label{fig:LiHoF4}
\end{figure}

In 1996 Bitko, Rosenbaum and Aeppli \cite{BRA96} measured the magnetic
properties of LiHoF$_4$ as a function of temperature and a
magnetic field which was applied perpendicular to the preferred
spin orientation. The resulting phase diagram is shown
in Fig. \ref{fig:LiHoF4}.
In order to understand this phase diagram we now consider a
minimal microscopic model for the relevant magnetic degrees of
freedom in LiHoF$_4$, the Ising model in a transverse field.
Choosing the $z$-axis to be the Ising axis its Hamiltonian is given by
\begin{equation}
H = -J \sum_{\langle ij \rangle}S_i^z \, S_j^z - h \sum_i S_i^x~.
\end{equation}
Here $S_i^z$ and $S_i^x$ are the $z$ and $x$ components of the Holmium spin
at lattice site $i$,
respectively. The first term in the Hamiltonian
describes the ferromagnetic interaction between the spins
which we restrict to nearest neighbors for simplicity. The second term is
the transversal magnetic field. For zero field $h=0$, the model reduces
to the well-known classical Ising model. At zero temperature all spins are
parallel. At a small but finite temperature a few spins will flip into the
opposite direction. With increasing temperature the number and size of the
flipped regions increases, reducing the total magnetization. At the critical
temperature (about 1.5 K for LiHoF$_4$) the magnetization vanishes, and
the system becomes paramagnetic. The resulting transition is a continuous
classical phase transition caused by thermal fluctuations.

Let us now consider the influence of the transverse magnetic field.
To do so, it is convenient to rewrite the field term as
\begin{equation}
-h \sum_i S_i^x = -h \sum_i (S_i^+ + S_i^-)
\end{equation}
where $S_i^+$ and $S_i^-$ are the spin flip operators at site $i$.
From this representation it is easy to see that the transverse
field will cause spin flips. These flips are the quantum fluctuations
discussed in the preceeding sections.  If the transverse field
becomes larger than some critical field
$h_c$ (about 50 kOe in LiHoF$_4$) they will destroy the
ferromagnetic long-range order
in the system even at zero temperature. This transition is a quantum phase
transition driven exclusively by quantum fluctuations.

For the transversal field Ising model the quantum-to-classical mapping
discussed in section \ref{sec:qm} can be easily demonstrated at a microscopic
level. Consider a one-dimensional classical Ising chain with the Hamiltonian
\begin{equation}
H_{cl} = -J \sum_{i=1}^N S_i^z \, S_{i+1}^z~.
\end{equation}
Its partition function is given by
\begin{equation}
Z= {\rm Tr}\, e^{-H_{cl}/T} = {\rm Tr}\, e^{J \sum_{i=1}^N S_i^z S_{i+1}^z / T}
   = {\rm Tr}\, \prod_{i=1}^N M_{i,i+1}~,
\end{equation}
where $\mathbf{M}$ is the so called transfer matrix. It can be represented as
\begin{equation}
\mathbf{M} = \left(
     \begin{array}{cc}
     e^{J/T} & e^{-J/T}\\ e^{-J/T} & e^{J/T}
     \end{array}\right)
     =e^{J/T} (1+ e^{-2J/T} S^x) \approx \exp(J/T + e^{-2J/T} S^x)~.
\end{equation}
Except for a multiplicative constant the partition function of the classical
Ising chain has the same form as that of a single quantum spin in a transverse
field, $H_Q= -h \, S^x$, which can be written as
\begin{equation}
Z= {\rm Tr} e^{-H_Q/T_Q} = {\rm Tr} e^{h S^x /T_Q} = {\rm Tr} \prod_{i=1}^N
   e^{h S^x /(T_Q N)}~.
\label{eq:quantum Z}
\end{equation}
Thus, a single quantum spin can be mapped onto a classical Ising chain. This
considerations can easily be generalized to a $d$-dimensional transversal field
(quantum) Ising model which can be mapped onto a $(d+1)$-dimensional classical
Ising model. Consequently, the dynamical
exponent $z$ must be equal to unity for the quantum phase transitions of
transverse field Ising models.
Using a path integral approach analogous to Feynman's treatment of
a single quantum particle, the quantum Landau-Ginzburg-Wilson functional
(\ref{eq:quantum LGW}) can be derived from (\ref{eq:quantum Z}) or its
higher-dimensional analogs.

\section{Quantum phase transitions and non-Fermi liquids}
\label{sec:non-fermi}

In this section we want to discuss a particularly important
consequence of quantum phase transitions, viz. non-Fermi liquid
behavior in an itinerant electron system.
In a normal metal the electrons form a Fermi liquid, a concept developed by
Landau in the 1950s \cite{Landau56}. In this state the strongly (via Coulomb potential)
interacting electrons essentially
behave like almost non-interacting quasiparticles with renormalized
parameters (like the effective mass). This permits very general and universal
predictions for the low-temperature properties of metallic electrons:
For sufficiently low temperatures the specific heat is supposed to be
linear in the temperature, the magnetic susceptibility approaches a constant,
and the electric resistivity has the form $\rho(T) = \rho_0 +A T^2$
(where $\rho_0$ is the residual resistance caused by impurities).

The Fermi liquid concept is extremely successful, it describes the
vast majority of conducting materials.
However, in the last years there have been experimental observations
that are in contradiction to the Fermi liquid picture, e.g.
in the normal phase of the high-$T_c$ superconducting materials
\cite{HighTc} or in heavy fermion systems \cite{HeavyF}. These are
compounds of rare-earth elements or actinides where the
quasiparticle effective mass is up to a few thousand times larger
than the electron mass.

\begin{figure}[t]
  \centerline{\includegraphics[width=9cm]{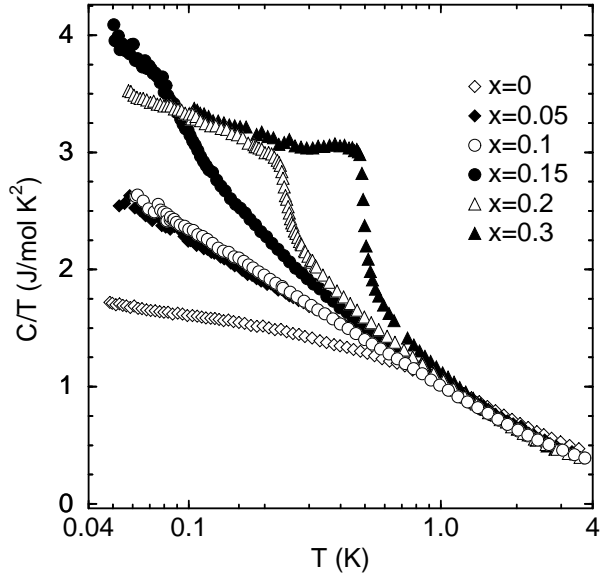}}
  \caption{Specific heat coefficient $C/T$ of CeCu$_{6-x}$Au$_x$
       as a function of temperature $T$ for different gold
           concentration $x$ (after \protect\cite{Lohn96}). }
\label{fig:specheat}
\end{figure}

Fig.\ \ref{fig:specheat} shows an example for such an observation,
 viz. the specific heat coefficient $C/T$ of the heavy-fermion system
von CeCu$_{6-x}$Au$_x$ as a function of the temperature $T$.
In a Fermi liquid $C/T$ should become constant for sufficiently
low temperatures. Instead, in Fig. \ref{fig:specheat} $C/T$ shows a pronounced
temperature dependence.
In particular, the sample with a gold concentration of $x=0.1$ shows a
logarithmic temperature dependence, $C/T \sim \log (1/T)$, over a wide
temperature range.

\begin{figure}[t]
  \centerline{\includegraphics[width=9cm]{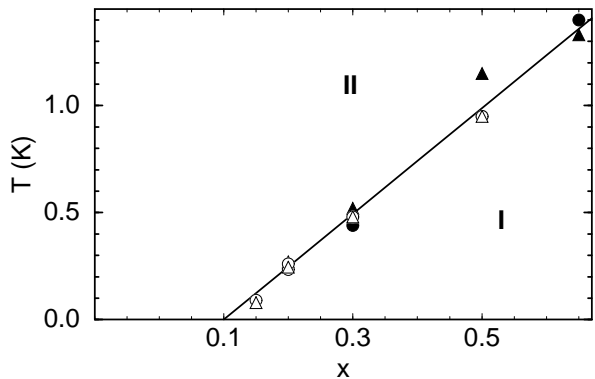}}
  \caption{Magnetic phase diagram of CeCu$_{6-x}$Au$_x$ (after \protect\cite{Lohn96}).
           I is the antiferromagnetic phase, II the paramagnetic phase.}
\label{fig:Cecuau}
\end{figure}

In order to understand these deviations from the Fermi liquid and
in particular the
qualitative differences between the behaviors at different $x$
it is helpful to relate the specific heat to the magnetic
phase diagram of CeCu$_{6-x}$Au$_x$, which is shown in Fig.\ \ref{fig:Cecuau}.
Pure CeCu$_6$ is a paramagnet, but by alloying with gold it can become
antiferromagnetic. The quantum phase transition is roughly at
a critical gold concentration of $x_c=0.1$ (and $T=0$).

Let us first discuss the specific heat at the critical concentration $x_c=0.1$.
A comparison with the schematic phase diagram in Fig.\
\ref{fig:schematic phase diagram} shows that at this concentration the entire
experiment is done in the quantum critical region (path (b)).
When approaching the quantum critical point, i.e. with decreasing temperature
the antiferromagnetic fluctuations diverge. The electrons are scattered off
these fluctuations which hinders their movement and therefore
increases the effective mass of the quasiparticles. In the limit
of zero temperature the effective mass diverges and with it the
specific heat coefficient $C/T$.

In an experiment at a gold concentration slightly above or below
the critical concentration the system will be in the quantum
critical region at high temperatures. Here the specific heat agrees
with that at the critical concentration.
However, with decreasing temperature the system leaves the quantum-critical
region, either towards the quantum disordered region (for $x<x_c$) or into
the ordered phase (for $x>x_c$).
In the first case there will be a crossover from the quantum critical behavior
$C/T \sim \log (1/T)$ to conventional Fermi liquid behavior
$C/T = {\rm const}$. This can be seen for the $x=0$ data in Fig.\
\ref{fig:specheat}.
In the opposite case, $x>x_c$ the system undergoes an
antiferromagnetic phase transition at some finite temperature,
connected with a singularity in the specific heat.
In Fig.\ \ref{fig:specheat} this singularity is manifest as a
pronounced shoulder.

In conclusion, the non-Fermi liquid behavior of  CeCu$_{6-x}$Au$_x$ can
be completely explained (at least qualitatively) by the antiferromagnetic
quantum phase transition at the critical gold concentration of $x_c=0.1$.
The deviations from the Fermi liquid occur in the quantum critical region
where the electrons are scattered off the diverging magnetic fluctuations.
Analogous considerations can be applied to other observables, e.g.
the magnetic susceptibility or the electric resistivity.

\section{Summary and Outlook}
Quantum phase transitions are a fascinating subject in todays condensed
matter physics. They open new ways of looking at complex situations and
materials for which conventional methods like perturbation theory fail.
So far, only the simplest, and the most obvious cases have been studied
in detail which leaves a lot of interesting research for the future.

This work was supported
in part by the DFG under grant Nos.
Vo659/2 and SFB393/C2 and by the NSF under grant No. DMR--98--70597.


%

\end{document}